%% file: vpn.tex
\documentclass[conference]{IEEEtran}
\makeatletter
\def\ps@headings{%
\def\@oddhead{\mbox{}\scriptsize\rightmark \hfil \thepage}%
\def\@evenhead{\scriptsize\thepage \hfil \leftmark\mbox{}}%
\def\@oddfoot{}%
\def\@evenfoot{}}
\makeatother
\IEEEoverridecommandlockouts
\usepackage{cite}
\usepackage{amsmath,amssymb,amsfonts}
\usepackage{algorithmic}
\usepackage{graphicx}
\usepackage{textcomp}
\usepackage{xcolor}
\def\BibTeX{{\rm B\kern-.05em{\sc i\kern-.025em b}\kern-.08em
    T\kern-.1667em\lower.7ex\hbox{E}\kern-.125emX}}

\usepackage[hidelinks,draft,bookmarks=false]{hyperref}

\usepackage{etoolbox}
\newtoggle{blinded}
\togglefalse{blinded}

\usepackage{caption}
\usepackage{subcaption}

\usepackage{placeins} 

\usepackage{stfloats}
\usepackage{float}

\usepackage{tabularx}
\usepackage{color, colortbl}

\usepackage{adjustbox}
\usepackage{array}
\newcolumntype{R}[2]{%
    >{\adjustbox{angle=#1,lap=\width-(#2)}\bgroup}%
    l%
    <{\egroup}%
}

\usepackage{wasysym}

\usepackage[binary-units]{siunitx}
\definecolor{darkred}{RGB}{212,0,16}
\definecolor{darkgreen}{RGB}{11,196,1}
\definecolor{darkergreen}{RGB}{6,98,1}
\definecolor{darkblue}{RGB}{0,73,218}

\definecolor{gerbertred}{RGB}{199,53,0}
\definecolor{gerbertgreen}{RGB}{130,179,102}
\definecolor{gerbertlightgreen}{RGB}{213,232,212}
\definecolor{gerbertgray}{RGB}{153,153,153}
\definecolor{gerbertorange}{RGB}{255,181,112}

\usepackage{listings} 
\lstdefinelanguage{ASM}{
    morekeywords={b, ble, blt, bne, bx, bl, ldr, str, push, pop, mov, add, sub},
    keywordstyle=\color{blue},
    sensitive=false, 
    morecomment=[l]{//}, 
    morecomment=[s]{/*}{*/}, 
    morestring=[b]", 
} %
\lstdefinelanguage{none}{
  identifierstyle=
}

\lstdefinelanguage{JavaScript}{
    morekeywords={function, var, return, while},
    keywordstyle=\color{gerbertred},
    keywords = [2]{openURL, confirm},
    keywordstyle=[2]\color{gerbertgreen},
    keywords = [3]{ObjC},
    keywordstyle=[3]\color{gerbertgray},
    sensitive=false, 
    morecomment=[l]{//}, 
    morecomment=[s]{/*}{*/}, 
    morestring=[b]", 
} %
\lstdefinelanguage{none}{
  identifierstyle=
}

\lstdefinelanguage{JSON}{
    morekeywords={AuxSettings, BasicSettings, VPNConfigurationId},
    keywordstyle=\color{gerbertgreen},
    keywords = [2]{0, 1},
    keywordstyle=[2]\color{gerbertred},
    keywords = [3]{UserGenerated, Certificate, university},
    keywordstyle=[3]\color{gerbertgray},
	stringstyle=\color{brown}\ttfamily,
    sensitive=false, 
    morecomment=[l]{//}, 
    morecomment=[s]{/*}{*/}, 
    morestring=[b]", 
} %

\usepackage{xspace}

\usepackage[nolist]{acronym}

\acrodefplural{PoC}[PoCs]{Proofs of Concept}

\begin{acronym}
\acro{SDR}{Software-Defined Radio}
\acro{AGC}{Automatic Gain Control}
\acro{GCI}{Global Coexistence Interface}
\acro{ECI}{Enhanced Coexistence Interface}
\acro{AFH}{Adaptive Frequency Hopping}
\acro{FEM}{Front-End Module}
\acro{SP3T}{Single Pole, Triple Throw}
\acro{HCI}{Host Controller Interface}
\acro{A2DP}{Advanced Audio Distribution Profile}
\acro{SCO}{Synchronous Connection-Oriented}
\acro{GPIO}{General Purpose Input Output}
\acro{ASLR}{Address Space Layout Randomization}
\acro{LMP}{Link Management Protocol}
\acro{LCP}{Link Control Protocol}
\acro{LNA}{Low-Noise Amplifier}
\acro{EIR}{Extended Inquiry Response}
\acro{CRC}{Cyclic Redundancy Check}
\acro{RTOS}{Real-Time Operating System}
\acro{QEMU}{Quick Emulator}
\acro{UART}{Universal Asynchronous Receiver Transmitter}
\acro{MitM}{Machine-in-the-Middle}
\acro{MAC}{Media Access Control}
\acro{BLE}{Bluetooth Low Energy}
\acro{ACL}{Asynchronous Connection-Less}
\acro{DSP}{Digital Signal Processing}
\acro{SDIO}{Secure Digital Input Output}

\acro{BCS}{Bluetooth Core Scheduler}
\acro{ELF}{Executable and Linking Format}
\acro{GIAC}{Global Inquiry Access Code}
\acro{JSON}{JavaScript Object Notation}
\acro{L2CAP}{Logical Link Control and Adaptation Protocol}
\acro{LMP}{Link Management Protocol}
\acro{PTM}{Pseudo Terminal Master}
\acro{PTS}{Pseudo Terminal Slave}
\acro{QEMU}{Quick Emulator}
\acro{RCE}{Remote Code Execution}
\acro{RFU}{Reserved for Future Use}
\acro{SVC}{Supervisor Call}
\acro{UART}{Universal Asynchronous Receiver Transmitter}
\acro{WICED}{Wireless Internet Connectivity for Embedded Devices}
\acro{MMIO}{Memory Mapped Input/Output}
\acro{LE}{Bluetooth Low Energy}
\acro{IoT}{Internet of Things}
\acro{IDE}{Integrated Development Environment}
\acro{ARM}{Advanced RISC Machine}
\acro{LM}{Link Manager}

\acro{RTOS}{Real-Time Operating System}
\acro{DMA}{Direct Memory Access}
\acro{RXDMA}{Receive Direct Memory Access}
\acro{NVRAM}{Non-Volatile Random-Access Memory}
\acro{ROM}{Read-Only Memory}
\acro{Rx}{Receive}
\acro{Tx}{Transmit}
\acro{SPI}{Serial Peripheral Interface}
\acro{MSP}{Main Stack Pointer}
\acro{PSP}{Process Stack Pointer}
\acro{RF}{Radio Frequency}
\acro{BLOB}{Binary Large Object}
\acro{SCO}{Synchronous Connection Oriented}
\acro{UAF}{Use-After-Free}
\acro{FHS}{Frequency Hop Sync}
\acro{CRC}{Cyclic Redundancy Check}
\acro{PDU}{Protocol Data Unit}
\acro{NFC}{Near Field Communication}
\acro{JPEG}{Joint Photographic Experts Group}
\acro{LPE}{Local Privilege Escalation}
\acro{DEP}{Data Execution Prevention}
\acro{XN}{eXecute Never}
\acro{LCP}{Link Control Protocol}
\acro{GATT}{Generic Attribute}
\acro{PoC}{Proof of Concept}
\acro{EDR}{Enhanced Data Rate}
\acro{MWS}{Mobile Wireless Standards}
\acro{HAL}{Hardware Abstraction Layer}
\acro{LR}{Link Register}
\acro{eSIM}{embedded-SIM}
\acro{RSP}{Remote SIM Provisioning}
\acro{PC}{Program Counter}
\acro{MD}{More Data}
\acro{LLID}{Logical Link Identifier}
\acro{SFI}{Serial Flash Interface}
\acro{EEPROM}{Electrically Erasable Programmable Read-Only Memory}
\acro{TOFU}{Trust On First Use}
\acro{CVE}{Common Vulnerabilities and Exposure}
\acro{PCIe}{Peripheral Component Interconnect Express}
\acro{SECI}{Serial Enhanced Coexistence Interface}
\acro{HID}{Human Interface Device}
\acro{DoS}{Denial of Service}
\acro{PTA}{Packet Traffic Arbitration}
\acro{AWMA}{Alternating Wireless Medium Access}
\acro{AP}{Access Point}
\acro{ECDH}{Elliptic-Curve Diffie--Hellman}
\acro{BT}{Bluetooth Classic}
\acro{SSP}{Secure Simple Pairing}
\acro{SC}{Secure Connections}
\acro{MIC}{Message Integrity Check}

\acro{VPN}{Virtual Private Network}
\acro{IPC}{Inter-Process Communication}
\acro{TLS}{Transport Layer Security}
\acro{ASA}{Adaptive Security Appliance}
\acro{TLV}{Type Length Value}
\acro{SNAK}{System Network Abstraction Kit}

\end{acronym}

\usepackage[colorinlistoftodos,prependcaption,disable]{todonotes} 
\presetkeys%
    {todonotes}%
    {inline,backgroundcolor=orange}{}
    
\usepackage{blindtext}

\usepackage{mdframed}

\begin{document}

\title{Very Pwnable Network: \\Cisco AnyConnect Security Analysis}

\iftoggle{blinded}{
\author{Authors removed for submission.}
}
{
\author{\IEEEauthorblockN{Gerbert Roitburd}
\IEEEauthorblockA{\textit{SEEMOO, TU Darmsdtadt} \\
groitburd@seemoo.de}
\and
\IEEEauthorblockN{Matthias Ortmann}
\IEEEauthorblockA{\textit{SEEMOO, TU Darmsdtadt} \\
mortmann@seemoo.de}
\and
\IEEEauthorblockN{Matthias Hollick}
\IEEEauthorblockA{\textit{SEEMOO, TU Darmsdtadt} \\
mhollick@seemoo.de}
\and
\IEEEauthorblockN{Jiska Classen}
\IEEEauthorblockA{\textit{SEEMOO, TU Darmsdtadt} \\
jclassen@seemoo.de}
}
}

\maketitle

\input{sections/abstract.tex}


\input{sections/introduction.tex}

\input{sections/related.tex}
\input{sections/background.tex}

\input{sections/design.tex}

\input{sections/fuzzing.tex}
\input{sections/ios.tex}

\input{sections/conclusion.tex}

\section*{Acknowledgment}

We thank the \emph{Cisco} incident response team for their timely answers
and 90-day disclosure coordination. Moreover, we thank \emph{Apple} for
confirming the crash behavior of \acp{VPN} that do not implement the
\emph{Always On VPN} feature and fixing network extension related vulnerabilities.

\iftoggle{blinded}{Further acknowledgments are blinded for review.}
{This work has been funded by the German Federal
Ministry of Education and Research and the Hessen State Ministry for
Higher Education, Research and the Arts within their joint support of
the National Research Center for Applied Cybersecurity ATHENE.}

\bibliographystyle{IEEEtran}
\bibliography{bibliographies}

\end{document}

%% file: sections/abstract.tex

\begin{abstract}

Corporate \acp{VPN} enable users to work from home or while traveling.
At the same time, VPNs are tied to a company’s network infrastructure, forcing users to install
proprietary clients for network compatibility reasons. VPN clients run with high privileges to encrypt
 and reroute network traffic. Thus, bugs in VPN clients pose a substantial risk to their users and in turn the corporate network. 
\emph{Cisco}, the dominating vendor of enterprise network hardware,
offers VPN connectivity with their \emph{AnyConnect} client for
desktop and mobile devices. While past security research primarily focused on the
\emph{AnyConnect Windows} client, we show that \emph{Linux} and \emph{iOS} are based
on different architectures and have distinct security issues.
Our reverse engineering as well as the follow-up design analysis and fuzzing reveal 13 new vulnerabilities.
Seven of these are located in the \emph{Linux} client.
The root cause for privilege escalations on \emph{Linux} is anchored
so deep in the client’s architecture that it only got patched with a partial workaround.
A similar analysis on \emph{iOS} uncovers three \emph{AnyConnect}-specific bugs
as well as three general issues in \emph{iOS} network extensions, which apply to all
kinds of VPNs and are not restricted to \emph{AnyConnect}.
\end{abstract}

%% file: sections/introduction.tex

\section{Introduction}





When corporations build an internal network, they often stick to the same
vendor for all components due to compatibility reasons. A vendor should offer
a variety of solutions meeting all the customer's needs. Creating and maintaining
such a product range is a huge effort, and, thus, the corporate network landscape
is dominated by very few vendors. 
\emph{Cisco}'s market share including \acp{VPN} and other enterprise network equipment is around \SI{50}{\percent}~\cite{ciscoanyconnectshare:2021}. 
Thus, users connecting to corporate \acp{VPN} will likely face a setup that
requires them to install the \emph{Cisco AnyConnect} client.
This client supports the most popular desktop and mobile operating systems
\emph{Windows}, \emph{Linux}, \emph{macOS}, \emph{iOS}, and \emph{Android}.
While they have platform-dependent feature sets, they are
all compatible with \emph{Cisco}'s \ac{ASA}, which, amongst others, also
provides \ac{VPN} server functionality.
As a product that is meant to provide secure network access and protect
corporate networks, \ac{VPN} clients should have a high security standard.

In this paper, we analyze the \emph{AnyConnect} client for \emph{iOS} and \emph{Linux}.
These operating systems have very different security mechanisms and network stacks,
enforcing a fundamentally different implementation on both platforms. We reverse-engineer the proprietary
clients and their operating system integration, analyze design issues, 
and test interesting interfaces with automated fuzzing.
On \emph{Linux}, issues anchored deep
in the client's architecture lead to privilege escalations that can only be prevented with
workarounds. Even after our report and an official advisory by \emph{Cisco}, the default configuration remains insecure.
On \emph{iOS}, our findings are not limited to the \emph{AnyConnect} client. 
Third-party \ac{VPN} applications can be integrated into the \emph{iOS} network
stack using the network extension framework, in which we uncover three issues.
Despite these findings, \emph{iOS} network extensions conceptually prevent a multitude of attack vectors that have been previously
reported for desktop clients.
Our main contributions are as follows.
\begin{itemize}
\item Reverse-engineering of the \emph{AnyConnect iOS} and \emph{Linux} client functionality
      to understand the underlying design and security assumptions.
\item Protocol and design analysis of the \emph{Linux} client, revealing one version
      downgrade and two privilege escalation bugs.
\item Analysis of the \emph{iOS} client, uncovering
      multiple issues, including plaintext data transmission.
\item Fuzzing of interfaces identified during the initial analysis, discovering
      multiple memory corruption bugs on \emph{Linux}, as well as a permanent \mbox{Wi-Fi}
      \ac{DoS} that persists through \emph{iOS} reboots.
\item Analysis under unstable networking conditions, revealing a double-free memory corruption bug in \emph{iOS}
      network extensions that can be triggered over-the-air.
\end{itemize}

We responsibly disclosed all identified issues.
The remainder of this paper is structured as follows.
Previous work is categorized to get a better understanding about which types of bugs affected which
components in~\autoref{sec:related}.
The security analysis in
\autoref{sec:background} focuses on the \emph{Linux} client while \autoref{sec:ios} focuses on the \emph{iOS}
client. Both sections follow the same structure that explains the client's design, the resulting security
assumptions as well as the individual findings.
\autoref{sec:conclusion} concludes this paper.

%% file: sections/related.tex

\section{Previous Work}
\label{sec:related}

An overview of security issues that existed prior to our work is shown in
\autoref{fig:vulnerabilityCategories}.
Since the first public release of an \emph{AnyConnect} vulnerability in 2011,
twelve vulnerabilities per year were published on average~\cite{cisco:securityAdvisories:2020}. 
\emph{Cisco} rates issues as \emph{low}, \emph{medium}, \emph{high}, or \emph{critical}.
However, no issue with the rating \emph{critical} was published since a remote code execution vulnerability in 2012.
We categorize all issues from \emph{Cisco's} security bulletins to get a better understanding before starting our own security analysis.

\definecolor{piedarkgray}{RGB}{77,77,77}
\definecolor{pielightorange}{RGB}{255,230,204}
\definecolor{pieblue}{RGB}{218,232,252}
\definecolor{piered}{RGB}{234,107,102}
\definecolor{piegreen}{RGB}{102,171,159}
\definecolor{piepink}{RGB}{248,206,204}


\begin{figure}[!b]
    \centering

\begin{tikzpicture}[minimum height=0.55cm, scale=0.8, every node/.style={scale=0.8}, node distance=0.7cm, font={\footnotesize}] 
\filldraw[draw=piedarkgray,fill=gerbertlightgreen] (0,0) -- (90:3.2cm) arc[start angle=90,end angle=318.5,radius=3.2cm] -- cycle;
\filldraw[draw=piedarkgray,fill=pielightorange] (0,0) -- (318.5:3.2cm) arc[start angle=318.5,end angle=374.9,radius=3.2cm] -- cycle;
\filldraw[draw=piedarkgray,fill=gerbertgray] (0,0) -- (374.9:3.2cm) arc[start angle=374.9,end angle=390.5,radius=3.2cm] -- cycle;
\filldraw[draw=piedarkgray,fill=pieblue] (0,0) -- (390.5:3.2cm) arc[start angle=390.5,end angle=403,radius=3.2cm] -- cycle;
\filldraw[draw=piedarkgray,fill=gerbertorange] (0,0) -- (403:3.2cm) arc[start angle=403,end angle=412.4,radius=3.2cm] -- cycle;
\filldraw[draw=piedarkgray,fill=piered] (0,0) -- (412.4:3.2cm) arc[start angle=412.4,end angle=418.7,radius=3.2cm] -- cycle;
\filldraw[draw=piedarkgray,fill=piegreen] (0,0) -- (418.7:3.2cm) arc[start angle=418.7,end angle=424.95,radius=3.2cm] -- cycle;
\filldraw[draw=piedarkgray,fill=piepink] (0,0) -- (424.95:3.2cm) arc[start angle=424.95,end angle=450,radius=3.2cm] -- cycle;


\fill[fill=piedarkgray] (300:3cm) circle (0.08cm);
\path[-,color=piedarkgray,thick] (300:3cm) edge node[sloped,yshift=0cm,align=right,text width=6cm] {\textcolor{black}{Cryptography}\\ 63.5\,\%} ++(6.2cm,0);

\fill[fill=piedarkgray] (335:3cm) circle (0.08cm);
\path[-,color=piedarkgray,thick] (335:3cm) edge node[sloped,yshift=0cm,align=right,text width=4.8cm] {\textcolor{black}{Privilege Escalation}\\ 15.7\,\%} ++(5cm,0);

\fill[fill=piedarkgray] (382:3cm) circle (0.08cm);
\path[-,color=piedarkgray,thick] (382:3cm) edge ++(0.7cm,-1.4cm);
\path[-,color=piedarkgray,thick] (382:3cm)++(0.7cm,-1.4cm) edge node[sloped,yshift=0cm,align=right,text width=4cm] {\textcolor{black}{Remote Code Execution}\\ 4.3\,\%} ++(4.2cm,0);

\fill[fill=piedarkgray] (396:3cm) circle (0.08cm);
\path[-,color=piedarkgray,thick] (396:3cm) edge ++(1.07cm,-1.2cm);
\path[-,color=piedarkgray,thick] (396:3cm)++(1.07cm,-1.2cm) edge node[sloped,yshift=0cm,align=right,text width=4cm] {\textcolor{black}{Denial of Service}\\ 3.5\,\%} ++(4.2cm,0);

\fill[fill=piedarkgray] (407.5:3cm) circle (0.08cm);
\path[-,color=piedarkgray,thick] (407.5:3cm) edge ++(1.45cm,-0.9cm);
\path[-,color=piedarkgray,thick] (407.5:3cm)++(1.45cm,-0.9cm) edge node[sloped,yshift=0cm,align=right,text width=4cm] {\textcolor{black}{Sensitive Information}\\ 2.6\,\%} ++(4.2cm,0);

\fill[fill=piedarkgray] (415.5:3cm) circle (0.08cm);
\path[-,color=piedarkgray,thick] (415.5:3cm) edge ++(1.8cm,-0.35cm);
\path[-,color=piedarkgray,thick] (415.5:3cm)++(1.8cm,-0.35cm) edge node[sloped,yshift=0cm,align=right,text width=4cm] {\textcolor{black}{Version Downgrade}\\ 1.7\,\%} ++(4.2cm,0);

\fill[fill=piedarkgray] (421.6:3cm) circle (0.08cm);
\path[-,color=piedarkgray,thick] (421.6:3cm) edge ++(2.08cm,0.25cm);
\path[-,color=piedarkgray,thick] (421.6:3cm)++(2.08cm,0.25cm) edge node[sloped,yshift=0cm,align=right,text width=4cm] {\textcolor{black}{Overflow}\\ 1.7\,\%} ++(4.2cm,0);

\fill[fill=piedarkgray] (437.5:3cm) circle (0.08cm);
\path[-,color=piedarkgray,thick] (437.5:3cm) edge ++(2.85cm,0.75cm);
\path[-,color=piedarkgray,thick] (437.5:3cm)++(2.85cm,0.75cm) edge node[sloped,yshift=0cm,align=right,text width=4cm] {\textcolor{black}{Other}\\ 7.0\,\%} ++(4.2cm,0);

\filldraw[draw=piedarkgray,fill=white] (0,0) circle (1.3cm);
\end{tikzpicture}

    \caption{Public security bulletin vulnerability categories.}
    \label{fig:vulnerabilityCategories}
\end{figure}
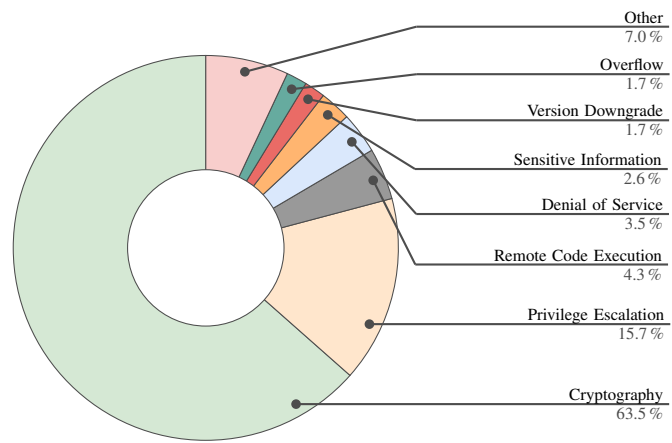

\subsection{Cryptography}
Not all published issues are directly located within the \emph{AnyConnect} client---the majority
is attributable to the third-party cryptographic library \emph{OpenSSL}.
This library had vulnerabilities like \emph{Logjam} and \emph{SWEET32}~\cite{cisco:CVE_2015_1788_Advisory:2015,cisco:CVE_2016_2177_Advisory:2016}. 
Thus, \SI{63.5}{\percent} of the issues fall into the cryptography category.
When analyzing \emph{Cisco}-specific issues, this category is out of scope.

\subsection{Privilege Escalation}
A \ac{VPN} client configures network routes and encrypts all
traffic. Hence, \emph{Any\-Connect} has components like \texttt{vpnagentd}
requiring system or administrator permissions. These pose an interesting
attack surface for privilege escalations. When excluding third-party components,
privilege escalations are the most frequently reported attack vector.
Since privilege escalations are systemic to \ac{VPN} clients and stem from
architectural issues, we provide further details about these.
However, the official advisories are missing this information, and we need
to rely on externally published write-ups.

One of the first privilege escalation vulnerabilities was published by Kostya
Kortchinsky in 2015~\cite{kortchinsky:AnyConnectEoP:2015}.
This vulnerability marks a turning point in identifying further, similar privilege
escalations. On \emph{Windows}, \path{vpnagent.exe} runs with system privileges.
Moreover, \path{vpnagent.exe} parses a special \ac{IPC} message that defines
a binary and arguments to execute it. This is not directly exploitable because
the binary needs to be signed by \emph{Cisco}. Kortchinsky identified a \emph{Cisco}-signed
executable \texttt{VACon64.exe}, which allows installing additional services, leading
to arbitrary code execution. 
\emph{Cisco} fixed this vulnerability by restricting \path{vpnagent.exe} to only
execute \path{vpndownloader.exe}.

Shortly after, James Forshaw and Yorick Koster independently discovered that this restriction did not include the full
path. An unprivileged user could copy \path{vpndownloader.exe}
to another directory and
plant a malicious \path{.dll} into the same directory, again leading to code execution~\cite{projectZero:AnyConnectPrivEsc:2015}.

In 2016, Duarte Silva found a flag in the same \ac{IPC} message~\cite{silva:AnyConnectPrivEsc:2016}.
This flag sets if \path{vpndownloader.exe} is launched from a temporary or secured application
directory. Unprivileged users can modify the temporary directory.

Koster's first report was a duplicate, but he identified a path
traversal issue several years later in 2020. The directory check was flawed because \emph{Windows} 
considers both \path{\} and \path{/} as directory separator~\cite{ssd:CVE_2020_3153:2020}.

All these issues are specific to \emph{Windows}. 
The history of these bugs shows that they were fixed individually but the
underlying design issues were not solved. Even the latest
advisory of January 2021 once again includes a malicious \path{.dll}
injection~\cite{dll:2021}.
To the best of our knowledge,
there are no write-ups for \emph{Linux} or \emph{iOS} privilege escalations. 

\subsection{Remote Code Execution}
All public vulnerabilities that lead to code execution require user interaction.
The only vulnerability with the rating \emph{critical}, fixed in 2012, still demands the user
to visit a malicious website that loads a \emph{Java} applet~\cite{cisco:historymultipleVulns:2012}. 

\subsection{Denial of Service}
The \ac{VPN} client might be interrupted or stopped. Depending on the remaining
system configuration, this can either lead to disrupted network connectivity or
traffic being exchanged without the additional layer of \ac{VPN} encryption.
For example, one vulnerability classified as \ac{DoS} allows a local attacker to stop
the \texttt{vpnagentd} service~\cite{cisco:historydosVuln:2020}.

\subsection{Sensitive Information}
A \ac{VPN} client has access to a lot of sensitive information that it could leak.
In a \ac{VPN} setting, remote information leakage is much more severe than local leakage.
One vulnerability of this category allows a remote attacker to exploit insufficient
boundary checks to read confidential system information~\cite{cisco:historyoobreadVuln:2019}.
This vulnerability is still only rated as \emph{medium}.

\subsection{Version Downgrade}
A version downgrade of the installed client is a first step to exploit previously fixed
and known vulnerabilities. The most severe issue in this category can be exploited remotely.
The web launch feature allows websites to start \emph{AnyConnect} using \emph{ActiveX} or
\emph{Java} applets, but it also allowed downgrading the \emph{AnyConnect} client~\cite{cisco:historydowngradeVuln:2012}.
Despite being exploitable remotely and an interesting component for exploit chains
that could lead to code execution, this bug was only rated as \emph{medium}.

\subsection{Overflow}
Missing input checks can cause buffer, heap, and integer overflows. These in turn lead
to crashes or might change the program flow.
Overflows could be assigned to the previous categories. However, they can be identified
automatically using fuzzing. From a security research perspective it is hence interesting
to list them separately to see how common simple programming mistakes are within the code base.
One of the overflow issues
allows an attacker to execute arbitrary code with system permissions~\cite{cisco:historybufferoverflowVuln:2013}.
It is only rated as \emph{medium}, similar to most other privilege escalation bugs.

\subsection{Other}
The remaining vulnerabilities stem from a variety of root causes.
Issues in this category include modifying configuration files as unprivileged user 
or damaging existing files owned by the system user.

%% file: sections/background.tex

\begin{table*}[!b]
\renewcommand{\arraystretch}{1.3}
\caption{Architectural and fuzzing issues in \emph{AnyConnect} for \emph{Linux}.}
\label{tab:vpnvulns}
\centering
\scriptsize
\begin{tabular}{|l|l|l|r|r|}
\hline
\textbf{Name} & \textbf{Cause} & \textbf{Impact} & \textbf{Report} & \textbf{Fix}\\
\hline
vpnagentd-vd & Missing version validation & Version downgrade & Jul 5 2020 & Sep 24 2020\\
vpnagentd-pe1 & Scripts can be overwritten & Vertical privilege escalation & Aug 6 2020 & Nov 4 2020*\\
vpnagentd-pe2 & Profiles can be overwritten & Vertical privilege escalation & Aug 6 2020 & Nov 4 2020\\
vpnagentd-c1 & Invalid memory address & Crash & Oct 28 2020 & Feb 24 2021\\
vpnagentd-c2 & Double free & Crash & Oct 28 2020 & Feb 24 2021 \\
vpnagentd-c3 & Heap corruption & Crash & Oct 28 2020 & Feb 24 2021 \\
vpnagentd-c4 & Heap corruption & Crash & Oct 28 2020 & Feb 24 2021 \\
\hline
\end{tabular}
\caption*{\normalfont\scriptsize{* Only a workaround configuration, still insecure by default.}}
\end{table*}

\section{Linux Client}
\label{sec:background}

\todo{ I would have been really interested by a more detailed analysis on this. Is it the same code base? Is it possible to perform binary diffing between the Linux and Windows versions? etc. -> answered parts of that but diffing is hard, different compiler options etc, meh}

In the following, we analyze the \emph{AnyConnect} client version 4.9.00086 on \emph{Linux}.
The overall architecture is similar to \emph{Windows}, and despite not having source-code access, we assume that major
parts of the code base are shared. 
However, both operating system have fundamentally different network stacks and different process
interaction, resulting in various implementation-specific details. Especially platform-dependent bugs
can therefore have similar root causes in the architecture, but require different fixes for each platform as they are not part of a shared code base.

We start with a component overview in \autoref{ssec:overview} and explain the basic connection setup in \autoref{ssec:connsetup}.
Based on this, we can make security assumptions in \autoref{ssec:assumptions}---for example,
the \ac{VPN} server must be ultimately trusted by a \emph{Linux} client to not execute malicious code. Even with this assumption,
the client's design allows for unpatchable privilege escalations, as explained in \autoref{ssec:design}.
Moreover, we identify further bugs with fuzzing in \autoref{ssec:fuzzing}.

\subsection{Component Overview}
\label{ssec:overview}
The \emph{Linux} client consists of three main binaries.

\begin{enumerate}
\item {\texttt{vpnagentd}} establishes \ac{VPN} tunnels and applies network settings.
\item {\texttt{vpnui}} is responsible for user interaction.
\item {\texttt{vpndownloader}} downloads profile files and updates provided by the \ac{VPN} server.
\end{enumerate}

The core \ac{VPN} functionality requires all these binaries and their interaction
via \ac{IPC}. \ac{IPC} messages are implemented as network messages sent to local TCP sockets.
The precise message format is \emph{Linux}-specific and shown in \autoref{tab:ipcmsg}.
While these messages are meant to enable communication between the three main binaries,
they also pose an attack surface to privilege escalations.

%

Moreover, the binaries rely on various libraries, including ports
of open-source libraries, as well as resources, which are processed
frequently and control its actions. The most noteworthy resources are profiles and local policies.
Profile files contain special features and rules to be used when connecting to a
specific \ac{VPN} server. The local policy file contains various settings also
affecting the security.
This overview is still very brief and simplified. The \texttt{/bin} directory
contains \num{8} binaries and \num{2} shell scripts, and \num{14} libraries reside
in the \path{/lib} directory.

\tikzset{>=latex}
\begin{table}[!b]
\vspace{-1em} 
\renewcommand{\arraystretch}{1.3}
\caption{IPC message format on \emph{Linux}.}
\label{tab:ipcmsg}
\centering
\scriptsize
\begin{tabular}{|l|l|l|}
 \hline
\textbf{Offset} & \textbf{Purpose} & \textbf{Default} \\ 
 \hline
\texttt{00-03}     & Magic byte         & OCSC \\ 
\texttt{04-05}     & Header length      & \texttt{26} \\
\texttt{06-07}     & Body length        &      \\
\texttt{08-0f}     & IPC response pointer  &      \\
\texttt{10-17}     & Unknown               &      \\
\texttt{18-1b}     & Unknown               &      \\
\texttt{1c-23}     & Return IPC object  &      \\
\texttt{24}        & Message type       &      \\
\texttt{25}        & Message identifier &      \\
\rowcolor{gerbertlightgreen}
\texttt{26-nn}     & Body               &      \\ 
 \hline
\end{tabular}\hspace*{0.5cm}\begin{tabular}{|l|l|}
\rowcolor{gerbertlightgreen}
 \hline
\textbf{Offset}    & \textbf{Purpose}       \\ 
 \hline
 \texttt{00-03}         & Type          \\ 
 \texttt{04-07}         & Length        \\
 \texttt{08-nn}         & Value         \\ 
 \hline
\end{tabular}

\vspace{-2.5cm}\hspace{2.7cm}\begin{tikzpicture}
    \path[->,gerbertgray] (0,0) edge [bend left=7] node [align=center,xshift=0.25cm,yshift=-0.5cm] {TLVs} (0.5,2.15);
	\end{tikzpicture}
	
\vspace{2em} 
\end{table}

\subsection{Connection Setup}
\label{ssec:connsetup}

The \emph{AnyConnect} client encrypts traffic based on \ac{TLS}.
Specific actions like authentication, file download, or tunnel setup
rely on HTTPS. The content-type of these messages is XML, which makes
interpretation rather simple.
Local \ac{IPC} uses plaintext communication in a binary format over TCP sockets.
On the server side, \emph{Cisco} \acp{VPN} run \ac{ASA}, which provides
remote access \acp{VPN}, site-to-site \acp{VPN}, and firewall functionality.

With all those binaries on the client side and configuration by
the \ac{ASA} server, the connection setup works as follows:

%
%

\begin{enumerate}
\item \texttt{vpnui} establishes a TLS connection to the \ac{ASA} server to perform
	  user authentication. The \ac{ASA} server requires a valid certificate and the user
	  has to provide valid credentials, meaning that both parties are mutually authenticated
	  after this step.
\item \ac{ASA} replies with an XML file, containing session tokens and a list of downloadable files.
\item \texttt{vpnui} launches \texttt{vpndownloader}, which continues running in the background.
\item \texttt{vpnui} transfers the essential XML contents to \texttt{vpndownloader} via IPC.
\item \texttt{vpndownloader} parses the XML and downloads available profile files and other resources.
\item \texttt{vpndownloader} notifies \texttt{vpnui} about the successful download.
\item \texttt{vpnui} advises \texttt{vpnagentd} to establish a \ac{VPN} tunnel.
\item \texttt{vpnagentd} sends a HTTPS \texttt{CONNECT} request to \ac{ASA} to initiate a tunnel.
\item \ac{ASA} replies with tunnel parameters such as a DNS server and routes.
\item From now on, \ac{VPN} traffic is exchanged between \texttt{vpnagentd} and \ac{ASA}.
\end{enumerate}

Even this simplified connection process shows the \emph{AnyConnect} complexity.
The binary separation and local communication via \ac{IPC} is meant to decouple processes running with user permissions
from \texttt{vpnagentd} with \texttt{root} privileges.

\subsection{Security Assumptions}
\label{ssec:assumptions}


We assume the \ac{ASA} \textbf{server has no malicious intent}. It could spy on the client's
network traffic or modify it. Moreover, the \texttt{vpndownloader} can be advised to download
scripts, executed by \texttt{vpnui}. Thus, the server needs to be
ultimately trusted. This is already quite exceptional, considering that the user does not get any
warnings if \ac{ASA} pushes scripts.
We set \texttt{vpndownloader} and \texttt{vpnui} out of scope since they do not run as privileged processes. 

Furthermore, we assume \textbf{\ac{TLS} is secure}. Breaking authenticated, end-to-end encrypted
communication has the same severe impact as a malicious server.
Moreover, the \textbf{server and client are mutually authenticated}. The server authenticates
with a certificate signed by a trusted authority and the client provides a certificate or credentials.

The client is using \textbf{Linux in default configuration} without special security mechanisms
activated. However, the operating system is compliant with a secure user role model. The \emph{AnyConnect}
client is also installed in default configuration, meaning that the application directory is readable
by all users but only writable with \texttt{root} privileges.

A local attacker aims at compromising confidentiality, integrity, and availability.
Moreover, they want to escalate their privileges.
They have the permission to run unprivileged code
in a shell and modify files in the \path{/tmp} directory. They can also open TCP ports above \num{1023} and
connect to TCP ports on the loopback interface.

%% file: sections/design.tex


\subsection{Architectural Issues and Logic Bugs}
\label{ssec:design}

Overall, we discovered three bugs by manually analyzing the protocol steps:
a version downgrade, overwriting scripts,
as well as overwriting profiles (see \autoref{tab:vpnvulns}).

\subsubsection{Version Downgrade}
The \path{vpndownloader} is responsible for downloading client updates from the
server. After a successful download, it sends an \ac{IPC} message specifying an installer executable.
Yet, the actual binary can
have an older version, and an attacker can replace it.
Prior to installation, the installer's hash and \emph{Cisco} signature
are verified, which prevents installing arbitrary software.
Nonetheless, it is possible to install any \emph{AnyConnect} version, including downgrades.

A downgrade enables exploitation of previously disclosed bugs. However, \emph{Cisco}
rated this version downgrade vulnerability so low that they did not publish any advisory. This is surprising given
that other downgrade vulnerabilities were included in advisories. Since version downgrades
have also been reported for other desktop clients, this indicates that \emph{Cisco}
does not validate and fix the root cause of each vulnerability in all clients.

\begin{figure}[!b]
    \centering
   \includegraphics[width=\columnwidth]{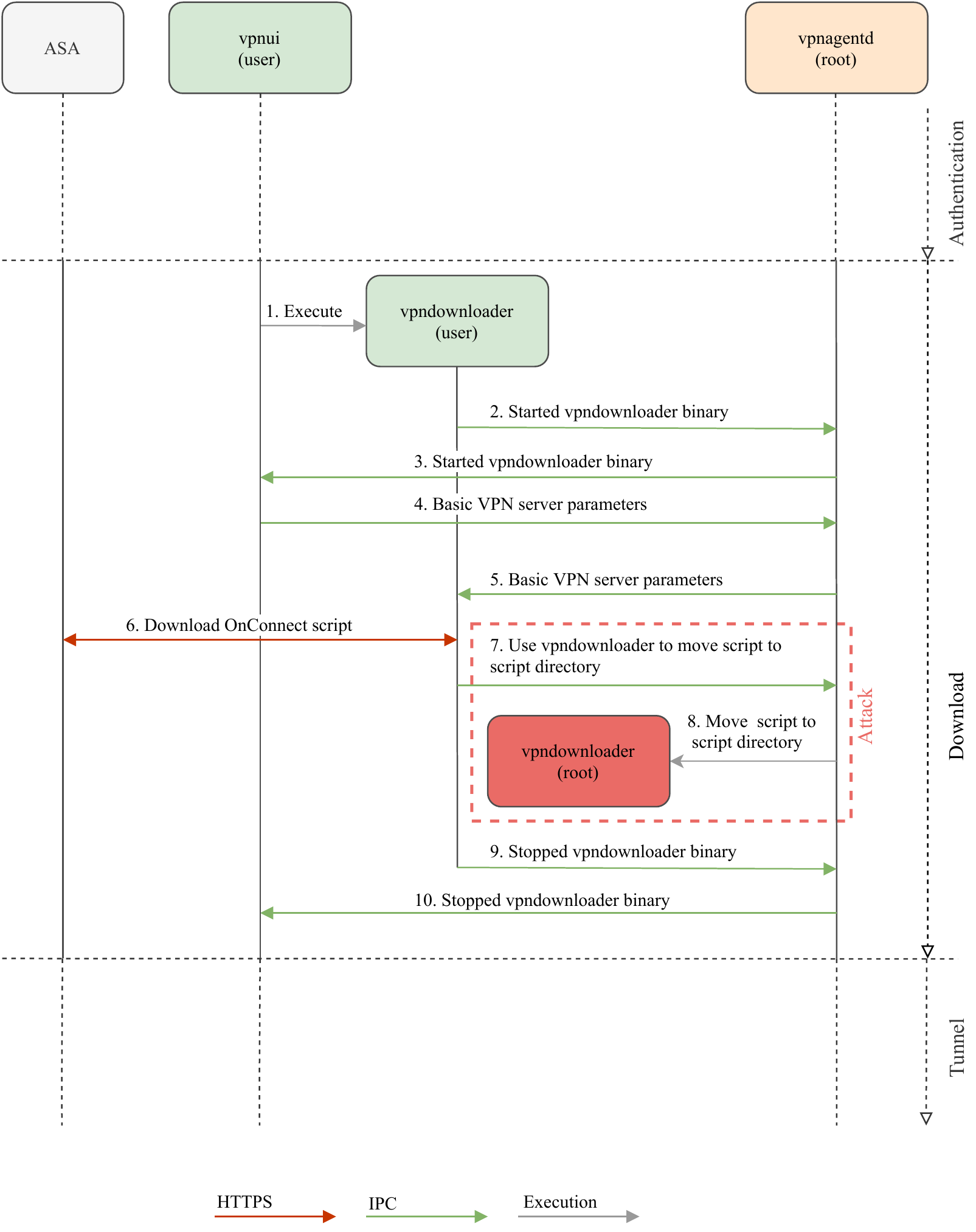} 
    \caption{Script deployment process.}
    \label{fig:basicOverviewScriptDeployment}
    \vspace{-0.5em} 
\end{figure}

\begin{figure}[!b]
\begin{lstlisting}[caption={Script deployment IPC message.},label=lst:ipcScriptDeployMessage]
@4f43 5343 2600 f400 ffff ffff ffff ffff@  @OCSC\&...........@
@0000 0000 0000 0000 0200 0000 0000 0000@  @................@
@0000 0000 0102 0001 0028 2f6f 7074 2f63@  @.........(/opt/c@
@6973 636f 2f61 6e79 636f 6e6e 6563 742f@  @isco/anyconnect/@
@6269 6e2f 7670 6e64 6f77 6e6c 6f61 6465@  @bin/vpndownloade@
@7200 0002 0094 \textcolor{gerbertgray}{22}\textcolor{gerbertred}{43 4143 2d6d 6f76 65}09@  @r.....\textcolor{gerbertgray}{"}\textcolor{gerbertred}{CAC-move}.@
@\textcolor{gerbertorange}{2d69 7063 3d33 3733 3139} 09\textcolor{gerbertgreen}{2f 746d 702f}@  @\textcolor{gerbertorange}{-ipc=37319}@.@\textcolor{gerbertgreen}{/tmp/}@
@\textcolor{gerbertgreen}{2e61 6348 314a 3333 422f 4f6e 436f 6e6e}@  @\textcolor{gerbertgreen}{.acH1J33B/OnConn}@
@\textcolor{gerbertgreen}{6563 745f 6c69 7474 6c65} 09\textcolor{gerbertgray}{2f 6f70 742f}@  @\textcolor{gerbertgreen}{ect\_little}@.@\textcolor{gerbertgray}{/opt/}@
@\textcolor{gerbertgray}{6369 7363 6f2f 616e 7963 6f6e 6e65 6374}@  @\textcolor{gerbertgray}{cisco/anyconnect}@
@\textcolor{gerbertgray}{2f73 6372 6970 742f 4f6e 436f 6e6e 6563}@  @\textcolor{gerbertgray}{/script/OnConnec}@
@\textcolor{gerbertgray}{745f 6c69 7474 6c65} 09\textcolor{gerbertred}{42 3446 4433 3833}@  @\textcolor{gerbertgray}{t\_little}.\textcolor{gerbertred}{B4FD383}@
@\textcolor{gerbertred}{3645 4338 3246 3146 3542 3544 3338 3437}@  @\textcolor{gerbertred}{6EC82F1F5B5D3847}@
@\textcolor{gerbertred}{4433 4132 4136 4142 3739 3032 4435 3438}@  @\textcolor{gerbertred}{D3A2A6AB7902D548}@
@\textcolor{gerbertred}{4209 7368 6131} 09\textcolor{gerbertgreen}{31} \textcolor{gerbertgray}{2200} 8005 0001 0006@  @\textcolor{gerbertred}{B}.\textcolor{gerbertred}{sha1}.\textcolor{gerbertgreen}{1}\textcolor{gerbertgray}{".}......@
@0028 2f6f 7074 2f63 6973 636f 2f61 6e79@  @.(/opt/cisco/any@
@636f 6e6e 6563 742f 6269 6e2f 7670 6e64@  @connect/bin/vpnd@
@6f77 6e6c 6f61 6465 7200@                 @ownloader.@
\end{lstlisting}
\vspace{-1em} 
\end{figure}

\subsubsection{Privilege Escalation}
An attacker can overwrite the \path{OnDisconnect} script and manually trigger it by
disconnecting from the \ac{VPN}. If scripting is not enabled in a profile, this
can be bypassed by also overwriting the profile. Both
can be combined for a reliable privilege escalation.

\paragraph{Overwriting Scripts}
Attackers can run scripts with permissions of active \ac{VPN} users,
since scripts are executed by \texttt{vpnui}. 
Vulnerable parts within the full script deployment process are depicted in \autoref{fig:basicOverviewScriptDeployment}.

During normal operation,  \texttt{vpndownloader}
stores scripts in a temporary directory and then advises the \texttt{vpnagentd} to move them to an \emph{AnyConnect} directory
using an \ac{IPC} message. This \ac{IPC} message contains a temporary script path, final script path, and script hash  as
shown in \autoref{lst:ipcScriptDeployMessage}. The \textcolor{gerbertred}{\texttt{CAC-move}} command takes both script paths
as argument. The hash value is used for a file integrity check and prevents \texttt{vpndownloader} to move files without
read access.
The value after the hash is set to \textcolor{gerbertgreen}{\texttt{1}}, meaning that the script will be saved with
\texttt{-rwxr-xr-x} permissions. 
Since this is an \ac{IPC} message from \texttt{vpndownloader} to \texttt{vpnagentd}, it also indicates its listening
port \textcolor{gerbertorange}{\texttt{37319}} for replies.
\texttt{vpnagentd} does not move the script directly but launches a second, privileged instance of \texttt{vpndownloader}
moving the scripts.

The \ac{IPC} messages lack authentication. Thus, every user on the system can send them to the \texttt{vpnagentd} port \num{29754}.
Moreover, all users can create scripts in the \textcolor{gerbertgreen}{\path{/tmp}} directory and they will be accepted by the \textcolor{gerbertred}{\texttt{CAC-move}} command that moves
them to the final directory.
An attacker can trigger \path{OnDisconnect} scripts immediately by sending an additional \ac{IPC} disconnect message.

\emph{AnyConnect} version 4.9.04053 adds a new configuration option as a workaround.
Using the \path{RestrictScriptWeb} \path{Deploy} 
 element in a local policy file, it is now possible
to skip the distribution of scripts. However, this is
set to \texttt{false} by default. Users need to know this specific setting and manually
disable it after the \emph{AnyConnect} client installation to prevent script deployment by
the server.

\paragraph{Overwriting Profiles}

In case a \ac{VPN} connection profile has scripting disabled, it is possible to activate scripting
by overwriting the profile.
The new profile needs to set the \texttt{EnableScripting} element to \texttt{true}.

The overall approach for overwriting profiles is similar to scripts.
Profiles are stored in XML format and non-executable.
Instead of setting the \ac{IPC} message's last value to 
\textcolor{gerbertgreen}{\texttt{1}}, it is set to \textcolor{gerbertgreen}{\texttt{0}},
which corresponds to \texttt{-rw-r--r--} permissions.

In contrast to scripts, profiles are usually only applied once, even when overwriting 
an existing profile. However, during a reconnect, \texttt{vpnui} reads and processes
the profile again. Under normal circumstances this would only be a simple bug that results
in unnecessary parsing overhead. In this scenario, a reconnect enables attackers apply
a new profile.


\emph{Cisco} treated both file override bugs as one, since the resulting code execution can optionally be prevented by
the new \path{RestrictScriptWebDeploy} flag. 
\textbf{The underlying bug that enables attackers to inject
arbitrary profiles and scripts remains unpatched and the default configuration is insecure.}
Most likely \emph{Cisco} decided to leave the script deployment intact by default due to the risk of
breaking existing setups. Leaving the \path{RestrictScriptWebDeploy} flag disabled by default
means that users need to manually enable this flag---also after every \emph{AnyConnect} client update.

%% file: sections/fuzzing.tex

\subsection{Inter-Process Communication Parsing Mistakes}
\label{ssec:fuzzing}

We further automate identifying bugs in the \emph{Linux} \ac{IPC} implementation
with fuzzing.
Based on the reverse-engineered \ac{IPC} message format, we can inject messages
while the \emph{AnyConnect} binaries are running and have Internet connectivity.
As listed in \autoref{tab:vpnvulns}, this reveals four individual bugs.

\subsubsection{Inter-Process Communication Message Format}


\emph{AnyConnect} implements \ac{IPC} on \emph{Linux} with TCP sockets.
The reverse-engineered message format is shown in \autoref{tab:ipcmsg}.
Similar to the message in \autoref{lst:ipcScriptDeployMessage}, all messages
start with the string \texttt{OCSC}. The messages can even have pointers
to objects or functions.
The body contains multiple \ac{TLV} fields, carrying the actual payload.

\todo{the whole fuzzing section is still a bit unsexy, idk... maybe make a fuzzer overview figure?}

\subsubsection{Fuzzing Setup}



We fuzz \texttt{vpnagentd}, because it runs with \texttt{root} privileges and parses \ac{IPC} messages.
The \texttt{vpnagentd} \ac{IPC} interface is a non-trivial target. Without source code, we can only perform
blackbox fuzzing. Additionally, \texttt{vpnagentd} requires a fully-functional network stack.
Moreover, some bugs might only
occur when injecting a message sequence or while \texttt{vpnagentd} is in a certain state. 
Thus, we fuzz on a fully functional \emph{Linux} system. In a first version of the fuzzer, we tried collecting coverage
with F\reflectbox{R}IDA~\cite{ole:frida:2020}. Coverage collection significantly slows down the target and
the resulting coverage is inconsistent
due to the target's statefulness. Instead, 
we create a dumb fuzzer that injects packets via a TCP socket. Once the socket is closed, 
\texttt{vpnagentd} likely crashed due to fuzzing. New inputs are generated with \emph{radamsa}~\cite{helin:radamsa:2020}.  

Our fuzzer discovers multiple memory corruption bugs
that require a sequence of packets.
Messages the fuzzer injects into \texttt{vpnagentd} can reach the whole network stack, and despite
not being state-aware, it is very stateful and logs packet sequences. If the
\texttt{vpnagentd} \ac{IPC} interface has been fuzzed before, this was likely only with a more common
single-packet, coverage-based fuzzer.

\subsubsection{Fuzzing Results}

In the following, we briefly describe the bugs found during fuzzing.

\todo{if space allows, add 1-2 of the message listings?}

\todo{The wording to describe some of the vulnerabilities left me confused about whether or not the vulnerability existed and what it is. For example, it's not clear to me how the Priv Esc in III.D.2 is a privilege escalation and I think the "Invalid memory address" may be an arbitrary pointer dereference or r/w, but it's not clear.}

\paragraph{Invalid Memory Address}

This bug requires sending multiple \ac{IPC} messages simultaneously to \texttt{vpnagentd}.  
If an \ac{IPC} message corresponds to a certain type (\texttt{type$>$0}) and ID (\texttt{id!=0xd||id!=0x00}), 
it is processed by the \path{IPCDepot}, which notifies all registered handlers. 
In one of the handlers, the \ac{IPC} message is then decomposed into its \ac{TLV} tuples.
\path{CSingleTLV::SetBuffer} is called to extract the value from a \ac{TLV} tuple
with a \path{memcpy} operation.
The application crashes when accessing the address pointer with a \texttt{SIGSEGV}.

\paragraph{Double Free}

The following bug can be triggered when replacing the message length in the header with zero.
This can lead to the message being rejected, which includes replacing the message's memory area with zeros.
Then, \path{operator.delete(this);} clears the address area belonging to the message.
This causes \texttt{free} being called twice for the same address range, resulting in a double free.
Calling \texttt{free} more than once to an address pointer damages the memory management data structure,
which can allow arbitrary memory writes. 

\paragraph{Heap Corruption \#1}

Another bug can be reproduced by extracting  an existing \ac{IPC} status message,
which contains a notification about the current download progress of \path{vpndownloader},
and  slightly modifying it.
This status message is sent by \texttt{vpndownloader} to \texttt{vpnagentd}, which forwards it to \texttt{vpnui}.
The \ac{TLV} tuple at offset \texttt{2e} defines the string to be displayed by \texttt{vpnui}.
Replacing the message's length field to \texttt{0006} before sending it to \path{vpnagentd} causes a crash.
However, the crash does not occur during message processing, since
it only corrupts parts of the \path{vpnagentd} heap. This results in a 
\texttt{SIGABRT} during later heap usage.
The crash is difficult to reproduce because it only occurs when a timer expires and a new network manager client object is created.
Several minutes can pass between sending the message and the \texttt{SIGABRT} signal.

\paragraph{Heap Corruption \#2}

Another discovered bug is based on the previous heap corruption.
The previous bug relied on a single \ac{IPC} message and waiting for timers to expire.
By sending a specially crafted message sequence, the bug can be triggered faster.
During the crash, a basic validation of the message takes place within \path{CIpcTransport::OnSocketReadComplete}.
This involves creating an empty response message stub, which in turn
calls \path{malloc}, and since the heap is already corrupted, this directly causes a \texttt{SIGABRT}.

\subsubsection{Bug Impact}
All \emph{AnyConnect} binaries are compiled with the most recent binary security features,
as tested with \texttt{checksec}~\cite{Alladoum:gef:2019}. The \texttt{checksec}
output looks as follows for all binaries:
\begin{lstlisting}[language=none]
@\textcolor{gerbertgreen}{gef>}@ checksec
@\textcolor{blue}{{[+]}}@ checksec for '/opt/cisco/anyconnect/bin/vpnagentd'
Canary                 : @\textcolor{gerbertgreen}{\checkmark}@ (value: 0x19c289dbfffe6c00)
NX, PIE, Fortify       : @\textcolor{gerbertgreen}{\checkmark}@ 
RelRO                  : @\textcolor{gerbertgreen}{Full}@ 
\end{lstlisting}
Thus, we were only able to crash \texttt{vpnagentd} but could not alter the
control flow. Note that advanced exploitation techniques might still allow
to exploit such bugs under certain conditions, and as such, they should be 
patched.

The \path{vpnagentd} service is managed by \path{systemd} and immediately
restarted upon a crash. Currently active \ac{VPN} connections are deactivated,
and connections residing on top of it might be dropped.
However, \emph{AnyConnect} cannot be used if \texttt{vpnagentd}
keeps crashing continuously. Thus, even simple crashes can be used for a
permanent \ac{DoS}, which might motivate the user to manually disconnect from the \ac{VPN}
and use a plaintext Internet connection.


%% file: sections/ios.tex

\begin{table*}[!b]
\renewcommand{\arraystretch}{1.3}
\caption{Bugs identified in \emph{AnyConnect} for \emph{iOS} and the \emph{iOS} network stack.}
\label{tab:vpnvulnsios}
\centering
\scriptsize
\begin{tabular}{|l|l|l|r|l|}
\hline
\textbf{Name} & \textbf{Cause} & \textbf{Impact} & \textbf{Report} & \textbf{Fix}\\
\hline
ios-plaintext & Missing crash handler & Data sent silently without VPN after a network extension crash & Dec 22 2019 & --- (won't fix) \\
ios-dos & Missing interface name validation & Permanent Wi-Fi DoS & Jan 28 2021 & iOS 14.6\\
ios-0click & Double-free when parsing configs & Network zero-click VPN crash with invalid memory access & Feb 3 2021 & iOS 14.6\\
anyconnect-crash1 & Memory corruption & Configuration string controlled memory access within \texttt{ACExtension} & Dec 13 2019 & Dec 18 2020* \\
anyconnect-crash2 & Memory corruption & Likely memory access in \texttt{ACExtension} & Dec 13 2019 & Dec 18 2020* \\
anyconnect-crash3 & Fixed dereference & Crash only & Dec 13 2019 & --- (non-reproducible) \\
\hline
\end{tabular}
\caption*{\normalfont\scriptsize{* Claimed to be patched by \emph{Cisco}, reproducibility of these bugs is limited.}}
\end{table*}

\section{\lowercase{i}OS Client}
\label{sec:ios}

The \emph{iOS} client implementation has a very different architecture
and feature set. As a result, the app components (see \autoref{ssec:componentsios}),
connection setup (see \autoref{ssec:connectionios}), and security assumptions (see \autoref{ssec:assumptionsios})
differ a lot.
Nonetheless, we identify multiple bugs
listed in \autoref{tab:vpnvulnsios}, which are explained in \autoref{ssec:fuzzingios} and
\autoref{ssec:crashesios}.
Based on these bugs, we start further manual analysis and find that an attacker that can drop or modify
network packets, such as disabling a Wi-Fi access point, can trivially cause VPN crashes without user interaction.
As explained in \autoref{ssec:0click}, these originate from a double-free memory access while parsing VPN configurations.


\begin{table}[!b]
\renewcommand{\arraystretch}{1.3}
\caption{\emph{AnyConnect} network extension modules on \emph{iOS}.}
\label{tab:iosfun}
\centering
\scriptsize
\begin{tabular}{|l|r|}
\hline
\textbf{Module} & \textbf{Number of Functions}\\
\hline
\texttt{AnyConnect} & \num{6653} \\
\texttt{ACExtension} & \num{13684} \\
\texttt{ACShareExtension} & \num{3557} \\
\texttt{ACSiriExtension} & \num{351} \\
\hline
\end{tabular}
\end{table}

\subsection{Component Overview}
\label{ssec:componentsios}
\emph{iOS} sandboxes all applications and limits system functions apps can access.
System functionality is provided via public frameworks, which abstract system functions
and add various checks. Creating \ac{VPN} connections
is part of the network extension framework~\cite{ios-networkextension}.
It offers multiple variants to integrate and implement \acp{VPN}.
%
The \ac{VPN} server component, \ac{ASA}, only supports
TLS. \emph{AnyConnect} must therefore use the \emph{Packet Tunnel Provider} feature of the network extension
framework. This is implemented in a custom network extension called \texttt{ACExtension}.
The extension encrypts traffic with the \emph{OpenSSL} library, similar to the \emph{Linux} implementation. 

The \emph{AnyConnect} application can be extracted and decrypted from a jailbroken \emph{iPhone} using
F\reflectbox{R}IDA~\cite{ole:frida:2020} for further analysis. The main app binary is called \texttt{AnyConnect}
and provides the user interface.
\ac{VPN} functionality is
contained in the \texttt{ACExtension} plugin. There are two more plugin components named \texttt{ACShareExtension} and
\texttt{ACSiriExtensionUI}. The main binary and plugins are non-stripped, they still contain symbol information and debug strings despite being a compiled binary.

The \emph{iOS} client only needs to implement a custom packet format on top of an existing \ac{VPN} interface.
Yet, the code base is gigantic.
\autoref{tab:iosfun} lists the number of functions per binary for the app version 4.9.00518,
which are \num{24245} in total. Major parts of the \ac{VPN} logic are shared with other platforms,
and only the necessary parts like the user interface and network extension are implemented in \emph{Objective-C}.
Despite the \emph{iOS} framework concept that should unify network extensions and encourage
light implementations, \emph{AnyConnect} on \emph{iOS} is very complex.

\subsection{Connection Setup}
\label{ssec:connectionios}
Setting up connections is based on the \emph{iOS} network extension, which creates a
tunnel interface to route traffic.
Outbound packets arriving on the tunnel interface are read by the network
extension, encapsulated, and sent to the \ac{VPN} server. The server unpacks
the packets and routes them to the final destination.
Similar, inbound packets from the server are encapsulated by the server, sent to the
client, unpacked by the network extension, and written to the tunnel interface.

Similar to the \emph{Linux} client, \emph{iOS} can apply profiles for a
connection. \emph{iOS} only implements a subset of the functions~\cite{ios-profiles}.
Due to \emph{iOS}-specific security restrictions, many features are impossible
to implement and will not have any effect when configured by the server.
However, there are also features specific to mobile clients, such
as the roaming behavior when switching between \mbox{Wi-Fi} and
Cellular. Moreover, rules for connect on demand can be configured,
which offers automatic \ac{VPN} connection establishment when detecting
pre-defined DNS names.

\subsection{Security Assumptions}
\label{ssec:assumptionsios}

The \emph{iOS} framework concept and application sandboxing protect users.
The most dangerous features like the
connect and disconnect scripts on desktop clients cannot be implemented by
\emph{iOS} apps. This narrows down attack vectors
for privilege escalations. Moreover, it means that an app user only needs to
\textbf{trust the \ac{VPN} server with their network traffic}. Replacing
traffic to exploit overflows within the client would still be
possible for someone controlling the server.
Compared to implementing scripting out-of-the-box, this
is a limited attack surface, and code injected this way would only run
in the context of the network extension.

Since \emph{iOS} apps can only be
updated via the official \emph{App Store} and updates are installed automatically,
\textbf{downgrade attacks via the app are prevented}. Users can still disable app updates
manually and run an outdated version. However, there is no \ac{VPN} client interface that
would allow downgrades or, worst case, installing arbitrary executables.

One feature provided by \emph{iOS} 
is the so-called \emph{Always On VPN}~\cite{ios-alwayson}. This feature ensures a \ac{VPN}
stays always activated, including across reboots. The only possibility to deactivate an
\emph{Always On VPN} is to uninstall the according \ac{VPN} profile in the settings menu.
\emph{AnyConnect} does not support this feature. Thus, once a \ac{VPN} connection is 
terminated, traffic is no longer tunneled through the \ac{VPN} and sent directly via
a potentially untrusted \mbox{Wi-Fi} without the additional TLS encryption layer.
Thus, if the \textbf{network extension crashes, traffic is sent without VPN encryption and rerouting}. 
While most apps should use TLS on top, some services and websites might be plaintext,
including DNS.
This is different from the \emph{Linux} configuration where \texttt{vpnagentd}
is automatically restarted by \texttt{systemd}. When the app crashes, it cannot
warn the user as it is already terminated, and \emph{iOS} does not warn the user either.
Upon our request \emph{Apple} confirmed that this is the
expected behavior. Since we believe that this is a dangerous and unexpected default behavior,
we list it as vulnerability \emph{ios-plaintext}.

%
%
%
%
%
%
%
%

\subsection{Fuzzing the Configuration Interface}
\label{ssec:fuzzingios}

The app can be almost completely controlled through a custom URL scheme
starting with \path{anyconnect://} followed by further action parameters~\cite{ios-urls}.
This includes creating connection entries, importing \ac{VPN} profiles, configuring localization, connecting with pre-filled credentials,
importing certificates, disconnecting from a \ac{VPN}, and closing the app.
Only non-destructive operations are possible via this custom URL scheme,
which follows \emph{Apple}'s recommendation for developers~\cite{ios-urls-guide}.
It is not possible to delete connection entries, profiles, or localizations.
However, profiles and localizations can be overwritten.
The supported actions are as follows.

\begin{itemize}
\item \texttt{create}: Create connection entries.
\item \texttt{connect}: Connect with a specific connection entry identified by its host.
\item \texttt{disconnect}: Terminate the current connection.
\item \texttt{close}: Dismiss the \emph{AnyConnect} user interface.
\item \texttt{import}: Import certificates, profiles, and localizations.
\end{itemize}

These actions can be customized with multiple parameters. For example,
to connect to a specific existing profile with pre-filled credentials
and opening a website in the \emph{AnyConnect} user interface
after a successful connection attempt, the following
parameters can be passed:

\begin{lstlisting}[language=none]
anyconnect://connect?host=vpn.example.com&prefill_username=user&prefill_password=password&onsuccess=http%3A%2F%2Fwww.example.com
\end{lstlisting}

Users need to manually enable this URL scheme via the setting \emph{External Control -- Enabled}.
Thus, this interface is neither controllable remotely nor a typical one-click attack.
Nonetheless, we can use it to automatically fuzz test the app's functions.
For this, we build a F\reflectbox{R}IDA-based fuzzer that opens \path{anyconnect://} URLs. 
%
Additionally, the user interface needs to be hooked with F\reflectbox{R}IDA
to automate the manual connection confirmation. The \path{AnyConnect} core module
that is responsible for the user interface implements user prompts, which we hook, is implemented in
\mbox{\path{CredentialPromptsViewController}.} 
We also automate further steps like deleting all created \ac{VPN} connections
later on.

%

Identified crashes are simple to verify by
providing the according URL via a browser.
Using this method, we found one bug in the \emph{iOS} network stack.
If the connection description string is too large, the \emph{iOS}-internal settings app becomes very slow and unresponsive.
Moreover, it is no longer possible to connect to \mbox{Wi-Fi} networks, even after a reboot. Thus, this is a \ac{DoS} that affects the whole
\emph{iOS} network stack. \emph{AnyConnect} needs to be uninstalled to get \mbox{Wi-Fi} working again. Sometimes, the
\emph{AnyConnect} \ac{VPN} profile is still cached, and reinstalling the app again leads to the same \mbox{Wi-Fi} \ac{DoS} even
without installing a profile.

\subsection{Regular Connectivity Issues and Crashes}
\label{ssec:crashesios}
As previously stated in \autoref{ssec:assumptionsios} and claimed as separate vulnerability \emph{ios-plaintext}, a network extension crash on \emph{iOS}
leads to plaintext network traffic being sent without warning the user.
These crashes occur frequently during regular usage.
Everything required to trigger these crashes is an unstable network connectivity
that switches between \mbox{Wi-Fi}, Cellular, and no connectivity. 

Nonetheless, bad network connectivity during regular usage yielded in three unique crashes
on the \emph{AnyConnect} versions 4.8.00825 and 4.8.01097.
\emph{Cisco} claims to have fixed two of these crashes. In fact, we were not able to reproduce
crashes with an up-to-date \emph{AnyConnect} client. However, reproducing a crash requires
physically moving to places with bad network connectivity. Moreover, the crashes only
occurred every few days to weeks prior to the claimed bugfix. We assume that \emph{Cisco}
was able to find the crash sources based on our reports and fixed them.

\subsubsection{Memory Corruption \# 1}

Upon a reconnect, the configuration is applied again and the tunnel is reinitialized.
This happens via the functions \path{PacketTunnelProvider_apply} \path{VpnConfig_cb} 
and
\path{-[PacketTunnelProvider initTunnelBuffers:]}. When calling \path{objc_release}
in the initialization function after calling \path{initWithCapacity}, a memory
corruption can occur. The accessed memory address is invalid because it contains
strings like \texttt{IPv4} and \texttt{86k\textbackslash n}, which likely stem from
the \ac{VPN} configuration. Thus, this memory access might be controllable by an attacker
that can modify network traffic or runs the server. This crash happened multiple times
with different configuration strings.

\subsubsection{Memory Corruption \# 2}

The second memory corruption causes a \texttt{SIGABRT} in \path{libsystem_malloc}
after calling the function \path{-[PacketTunnelProvider write} 
\path{Packet:dataLen:isIPv4:]}.
The full stack trace originates from the event handler when a TLS packet is received,
which in turn calls \path{CTlsProtocol::OnSocketReadComplete}. After further intermediate
function calls, this results in calling \path{CTunTapMgr::postHostBoundPacket}.
There is one more call to the \path{AxtSNAKTuntap::Write} handler before finally
crashing in \path{-[PacketTunnelProvider writePacket:dataLen:isIPv4:]}.
Most parts in the stack trace sound rather generic, except from the \emph{Cisco}-specific
\ac{SNAK}. Even though the crash only resulted in an abort instead of an invalid memory
access, crashing via \path{malloc} indicates a memory corruption. This crash
only occurred once.

\subsubsection{Fixed Dereference}

The third crash does not look controllable and is a simple fixed dereference at a
pointer to \path{0x04}. While it is not worth to reverse-engineer its origin to determine
if it could be controllable by an attacker, it is
still a crash that disconnects from the \ac{VPN} server.

\subsubsection{Crash Debugging}

The \emph{AnyConnect} app has a configuration option to enable debug logs. These
logs contain messages detailing how connections are established
and which settings are applied. After enabling debug logs on one of our test devices,
\emph{AnyConnect} kept crashing---but without producing \emph{iOS} crash
logs and without saving debug logs that lead to the crash. This crash behavior 
might explain why \emph{Cisco} was not able to identify such issues during internal
testing.

\subsection{Attacker-Controlled Connectivity Issues and Crashes}
\label{ssec:0click}

We were not able to
program a \emph{Packet Tunnel Provider} fuzzer that causes exactly same behavior as described in the previous section.
Switching network interfaces and reconnecting to \acp{VPN} are implemented within the kernel. 
Hence, this cannot be reproduced by injecting packets into the \emph{Packet Tunnel Provider}, which
is only responsible for encrypting and decrypting packets on an upper network layer before
forwarding them to the \emph{iOS} \ac{VPN} tunnel interface. As of now, F\reflectbox{R}IDA only works in the user space.
Fuzzing techniques that also apply to the kernel were published recently~\cite{p0ned}, but still come with
a lot of limitations like restriction to selected modules and a lot of customized harnessing.

Even without specialized tooling, multiple actions can shut down the network interface or drop packets during
\ac{VPN} connection establishment. For testing purposes, the \texttt{ifconfig} command can be used
on jailbroken \emph{iPhones}, but connections can also be interrupted as a regular user on a standard
device by switching off Wi-Fi via menus.
The same behavior can be achieved \textbf{without user interaction}
by manually switching of the Wi-Fi access point, which is something any attacker within wireless
range could do by jamming packets.

Interestingly, the network interface state change caused by all of the above options triggers
a completely new bug. While originally trying to reproduce the \emph{AnyConnect} bugs after
\emph{Cisco} claimed to have fixed them, this regularly causes another crash due to an invalid memory access
in the \emph{iOS} network extension agent process \texttt{neagent}.
The crash happens in the ~\path{com.apple.NSXPCConnection.user.} 
\path{endpoint} thread while deallocating
an immutable dictionary (\texttt{NSDictionaryI}) object. Memory corruptions during deallocation
are also known as double-free, which can be abused for accessing arbitrary memory. Memory control
depends a lot on the object causing the double-free. The initial crash log contains
\num{19} entries just for the backtrace of the crashed thread, plus various additional information
on register states. To locate the actual root-cause and freed object, we use the \path{frida-trace} tool
to print dictionary access in \texttt{neagent} observed in the initial crash log.
The trace output when interrupting \ac{VPN} connection establishment via Wi-Fi
looks as follows:

\begin{lstlisting}[language=C]
/* TID 0x26803 */
-[__NSDictionaryI dealloc]  // repeated 18 times
-[__NSXPCInterfaceProxy_NEVPNPluginDriver
    startWithConfig:0x10109c620 complHandler:0x16f4ce200]
-[__NSDictionaryI dealloc]
/* TID 0x1864b */
-[__NSXPCInterfaceProxy_NEVPNPluginDriver
    startWithConfig:0x0 complHandler:0x16f3b65c0]
-[__NSXPCInterfaceProxy_NEVPNPluginDriver
    startWithConfig:0x1010a8ae0 complHandler:0x1010a6c40]
-[__NSXPCInterfaceProxy_NEVPNPluginDriver
    startWithConfig:0x0 complHandler:0x16f3b6198]
/* TID 0x1954b */
-[NEConfiguration .cxx_destruct]
| -[NEVPN .cxx_destruct]
| | -[__NSDictionaryI dealloc]  @\textcolor{darkred}{// double-free happens here}@
\end{lstlisting}

F\reflectbox{R}IDA has a backtracing functionality that provides a similar output as the \emph{iOS}-internal
crash log format. Using this backtrace, we confirm that the call trace when reaching the last \texttt{NSDictionaryI} \texttt{dealloc}
call shown in the previous listing looks similar to the \num{19} entries in the original crash log.
The deallocated object can be determined by hooking the \texttt{dealloc} function on entry and iterating
through all passed arguments as follows:

\begin{lstlisting}[language=JavaScript]
var dict = new ObjC.Object(args[0]);
var enumerator = dict.keyEnumerator();
var key;
while ((key = enumerator.nextObject()) !== null) {
	var value = dict.objectForKey_(key);
}
\end{lstlisting}

Using this technique, the double-freed dictionary turns out to be the
network extension configuration of the \ac{VPN} profile, which is stored
as JSON and contains partially controllable contents.



%% file: sections/conclusion.tex

\section{Conclusion}
\label{sec:conclusion}

Corporate \ac{VPN} solutions cannot provide the security they promise,
if they continue to be developed in a non-security conscious fashion.
Ideally, they add encryption to facilitate secure access to corporate networks. At the same time,
their ultimate control over a user's network traffic and integrated scripting engines
controllable via the server significantly endanger an end-user's system security.
Similar to previous findings in \emph{Cisco AnyConnect} on \emph{Windows}, the desktop client for \emph{Linux}
has various possibilities for privilege escalations and allows the server to push
scripts to be executed on the client by default.
Our findings show that even the restricted \emph{iOS} network extension framework
allows integrating bloated \ac{VPN} clients and has severe bugs in itself.
Despite being marketed as security product, users should be very skeptical
about installing and using \ac{VPN} clients.